\documentclass[twocolumn,aps,epsfig,superscriptaddress,prl]{revtex4}
\usepackage{amsmath}
\usepackage{amsfonts}
\usepackage{amssymb}
\usepackage{graphicx}
\usepackage{graphicx}
\usepackage{dcolumn}
\usepackage{bm}
\usepackage[sort&compress]{natbib}
\bibpunct{[}{]}{,}{n}{,}{,}

\begin{document}
\preprint{APS/123-QED}

\title{Quasimonoenergetic electron beams produced by colliding cross-polarized laser pulses in underdense plasmas}

\author{C. Rechatin}

\author{J. Faure}
\affiliation{%
Laboratoire d'Optique Appliqu\'ee, ENSTA, CNRS, Ecole Polytechnique, UMR 7639, 91761 Palaiseau, France\\
}%
\author{A. Lifschitz}%
\affiliation{%
Laboratoire d'Optique Appliqu\'ee, ENSTA, CNRS, Ecole Polytechnique, UMR 7639, 91761 Palaiseau, France\\
}%
\affiliation{Laboratoire de Physique des Gaz et des Plasmas, CNRS, UMR 8578, Universit\'e Paris XI, B\^atiment 210, 91405 Orsay cedex, France\
}
\author{X. Davoine}
\author{E. Lefebvre}
\affiliation{%
D\'epartement de Physique Th\'eorique et Appliqu\'ee, CEA, DAM Ile-de-France, BP 12, 91680 Bruy\`eres-le-Ch\^atel, France\
}
\author{V. Malka}
\affiliation{%
Laboratoire d'Optique Appliqu\'ee, ENSTA, CNRS, Ecole Polytechnique, UMR 7639, 91761 Palaiseau, France\\
}%
\date{\today}

\begin{abstract}
The interaction of two laser pulses in an underdense plasma has
proven to be able to inject electrons in plasma waves, thus
providing a stable and tunable source of electrons. Whereas
previous works focused on the ``beatwave'' injection scheme in
which two lasers with the same polarization collide in a plasma,
this present letter studies the effect of polarization and more
specifically the interaction of two colliding cross-polarized
laser pulses. It is shown both theoretically and experimentally
that electrons can also be pre-accelerated and injected by the stochastic heating occurring at the collision of two cross-polarized lasers and thus, a new regime of optical injection is demonstrated. It is found that injection with cross-polarized lasers occurs at higher laser intensities.
\end{abstract}
\maketitle

Plasma-based accelerators can sustain high electric fields in
excess of 100 GV/m, which is approximately 1000 times beyond what
can be achieved in conventional accelerators. In the plasma
medium, an intense laser pulse can drive a longitudinal wave,
called wakefield, travelling with a phase velocity $v_p=v_g$,
where $v_g$ is the group velocity  of the laser, and can be
close to the speed of light $c$. This high phase velocity
enables ultra relativistic acceleration \cite{Tajima&Dawson}, but
also requires that the initial speed of the particles is high
enough to allow trapping in the plasma wave. Whereas in a linear
regime, an electron with no initial velocity is not trapped, in a
more nonlinear regime, transverse wave breaking effects\cite{Rosenzweig,Bulanov}  can result
in the self-trapping of electrons in the so-called ``bubble
regime'' \cite{Bubble}. In this case, the injection of electrons
occurs in a short space-time volume and leads to the production of
narrow energy spread electron beams as observed in Ref.
\cite{Nature1,Nature2,Nature3}. Nevertheless, in this scheme,
self-injection and acceleration depend on the precise evolution of
the laser pulse. Therefore, fine control over the output electron
beam is hard to achieve. The precise control of electron injection
can be achieved by all-optical external injection schemes. These
schemes use two pulses with the same polarization that will
collide either head on \cite{Esarey,Fubiani} or with
an angle \cite{Umstadter,Hemker}. The collision of the pulses heats the
electronic population and some of the electrons can be trapped.
Previous publications \cite{NatureCPI,PPCF} have shown that this
external injection scheme with two linearly polarized lasers
allows a fine control over the injection and therefore brings
tremendous flexibility to laser-plasma accelerators. In
particular, it makes it possible to tune the output charge and
energy of the electron beam in a stable way.

\par
In this injection scheme, the collision of the two laser fields
creates a slow ponderomotive beatwave $\displaystyle \phi_b
\propto < \bf{a_0 . a_1} >$ where $\bf{a_{0}}$ and $\bf{a_1}$ are
respectively the normalized vector potentials of the main and
injection pulses. For the circular polarizations case the motion of
electrons is integrable and if both lasers have the same frequency
$\omega_0$, the maximum energy gain is $2 \sqrt{a_0 a_1}$. This
preacceleration then allows the electrons to be trapped in the
main plasma wave and to be further accelerated. The ponderomotive
beatwave acceleration can be conversely seen as a heating process.
The motion of electrons in two parallel polarized laser fields is
not analytically tractable since there is not only the slow
ponderomotive beatwave but also a fast varying component at
$2\omega_0$. For a colliding (head-on) geometry and above a
certain threshold, the motion becomes stochastic
\cite{Mendonca,stoc}. In this case, the collision of the laser
pulses leads to an even more efficient heating of the electrons.
In the case of cross-polarized lasers $\bf{a_0 . a_1}=0 $ and
there is no ponderomotive beatwave, thus no beatwave injection.
However, we will show experimental and numerical
evidence that, under some conditions, injection of high quality
electron beams is still possible with cross-polarized laser
pulses and that their features are comparable with the beatwave injected beams.

\par
\begin{figure}[!ht]
\includegraphics[width=8.7cm]{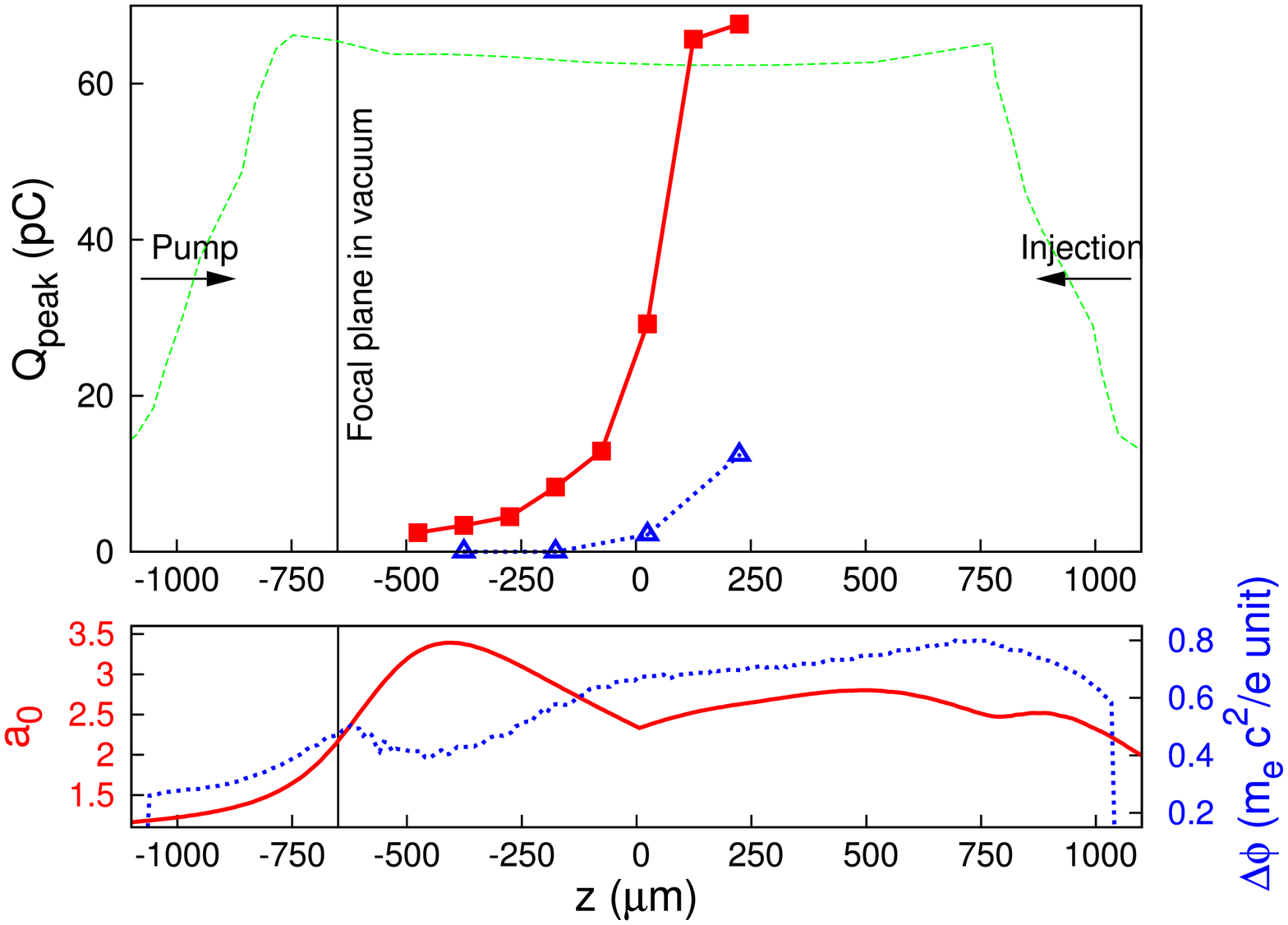}
\caption{\label{evol} Top: Charge of the monoenergetic peak for
parallel (solid squares line) and crossed (dotted triangles line)
polarizations. The measured plasma profile is also represented
(slash dotted line). Bottom: Simulated evolution of $a_0$ (solid
line) and $\Delta\phi$, average depth of the plasma wave potential
in the accelerating and focusing regions (dotted line).}
\end{figure}

The experiment was performed on the ``Salle Jaune'' laser system
at LOA. It delivers $720 \mbox{mJ}$ (pump pulse) and
$250\mbox{mJ}$ (injection pulse) at $\lambda_0=0.8 \;\mu\mbox{m}$
on target with Full Width Half Maximum (FWHM) duration $30
\mbox{fs}$. The pump pulse is focused by a 1 m focal length
spherical mirror to an intensity of $I_0=3.4\times\;10^{18}
\mbox{W.cm}^{-2}$ thus giving a normalized strength parameter
$a_0=1.3$. The injection pulse is focused by a 1 m focal length
off axis parabola to an intensity of $I_1=4\times\;10^{17}
\mbox{W.cm}^{-2}$, corresponding to $a_1=0.4$. The two collinear
and counter-propagating lasers are focused on the edge of a 2 mm
supersonic Helium gas jet whose density profile is represented in
figure \ref{evol}. The electronic density plateau corresponds to
$n_e=7.5\times\;10^{18} \mbox{cm}^{-3}$. The spectrum of electrons
is recorded by a LANEX screen coupled with a permanent magnet
\cite{aimantyannick}. It is set to record energies from 40 MeV to
400 MeV. It also gives access to the charge and divergence of the
electron beam. Finally, a half-wave plate makes it possible to
rotate the polarization of the pump pulse during the experiment,
thus allowing to study and compare the parallel and crossed
polarizations cases. The collision point of the two lasers could
be tuned simply by changing the time delay between the two pulses.
This is an important feature of the experiment since nonlinear
evolution of the laser in the plasma -due to self-focusing
\cite{Selffoctheory,Selffoc} and self-compression
\cite{Selfmodtheory,Selfmod}- changes the laser intensity along
the position in the gas jet. We used the particle code WAKE
\cite{Wake} to simulate the nonlinear evolution of the laser pulse
and the corresponding wakefield in cylindrical geometry. The
results of this simulation are shown in the bottom frame of figure
\ref{evol}: the solid line represents the evolution of $a_0$ and
the dotted line represents the average depth of the plasma wave potential
in the accelerating and focusing regions $\Delta
\phi$ (the average was made over $18
\;\mu\mbox{m}$, i.e. the focal spot FWHM). This parameter
characterizes the ability of the wakefield to trap preaccelerated
electrons since it is directly related to the minimum momentum
that an electron needs to be trapped, in normalized units: $
\displaystyle u_{z,min}
=\beta_p\gamma_p(1+\gamma_p\Delta\phi)-\gamma_p\sqrt{(1+\gamma_p
\Delta\phi)^2-1}$, where $\beta_p=v_p/c$ and
$\gamma_p=(1-\beta_p^2)^{-1/2}$. The nonlinear evolution of
the pulse allows us to indirectly obtain data over a broad range
of laser amplitudes $a_0$ at the collision of the two lasers and
thus to change the heating conditions. Simultaneously, the
wakefield becomes more suitable for trapping as the pump pulse
self-compresses and distorts spatially.

The experimental results are summarized in the top frame of figure
\ref{evol}. The graph represents the charge of the monoenergetic
component of the spectrum for different collision positions for
the cases of parallel and crossed polarizations. The figure shows
that in the case of parallel polarizations, injection of a
monoernergetic beam starts for $z_{inj}>-500\;\mu\mbox{m}$. This
corresponds to positions where the laser pulse has self-focused
(figure \ref{evol}, bottom). From that point, the charge increases
as injection occurs further inside the gas jet. The increase of
the charge can be explained by the evolution of the wakefield: as
the laser propagates in the gas jet, $\Delta\phi$ increases,
making the wakefield more suitable for trapping a large amount of
electrons (see also Ref. \cite{agustin} for more detailed
simulations). For the case of crossed polarizations, the behavior
is different: there is a threshold behavior and
monoenergetic electron beams are injected only for
$z_{inj}>25\;\mu\mbox{m}$. The charge also increases with $z$, as
in the parallel polarizations case, but it is lower by a factor of
5-6.

\begin{figure}[!ht]
\includegraphics[width=8.5cm]{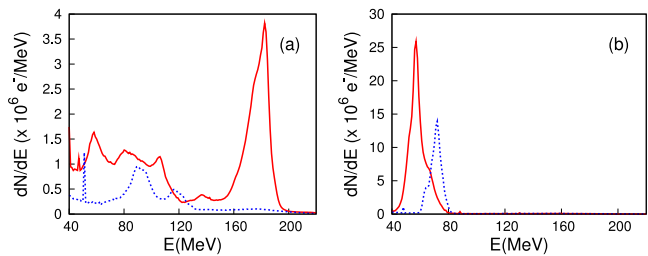}
\caption{\label{exp}Experimental spectra for
$z_{coll}=-175\;\mu\mbox{m}$ (a), $z_{coll}=225\;\mu\mbox{m}$ (b):
solid line: parallel polarizations, dotted line: crossed
polarizations.}
\end{figure}

Figure \ref{exp} shows typical electron spectra obtained with
parallel (solid line) and crossed (dotted line) polarizations, at
two injection positions (a: $z_{inj}=-175\;\mu\mbox{m}$ and b:
$z_{inj}=225\;\mu\mbox{m}$). At $z_{inj}=-175\;\mu\mbox{m}$, we
have observed a stable monoenergetic beam at 170 Mev with low
charge ($\simeq 10 \mbox{pC}$) in the parallel polarizations case.
When the polarizations are crossed, the high energy monoenergetic
component of the spectrum vanishes and a broad component at lower
energy remains.
When the collision takes place further inside the gas jet, e.g. at
$z_{inj}=225\;\mu\mbox{m}$, trapping become easier. Thus, for the
parallel polarizations case, a $50$ MeV beam with 68 pC is
injected. The decrease of energy from position (a) (170 MeV) to
(b) (50 MeV) can easily be explained by the decrease of the
remaining acceleration length after the injection
\cite{NatureCPI}. In the crossed polarizations case, the striking
result is the production of a stable quasi-monoenergetic electron
beam (see figure \ref{exp}b). The beam is stable in energy $E=72
\mbox{MeV} \pm 6 \mbox{MeV}$, in energy spread $\delta E/E = 14 \%
\pm 2 \% $ (FWHM), divergence $5 \pm 2\mbox{mrad}$ and charge
$Q=11 \pm 3 \mbox{pC}$. The injection of such a beam cannot be
explained by the beatwave scheme since the residual parallel
polarization component due to the possible half-wave plate misalignment is well below the threshold for beatwave
injection. Other heating mechanisms that are not polarization
dependent such as ponderomotive heating by the colliding envelopes
of the lasers \cite{Fubiani}, phasekick injection where a
colliding laser stimulates the wavebreaking of a nonlinear wake
\cite{Cary1}, or even wake collision \cite{Cary} could be possible
explanations. Here, we will show that the features of this
injection of electrons in a cross-polarized scheme can be
explained by the concurrence of two physical phenomena: the
heating of electrons during the laser collision and the plasma
wake inhibition.

First, we will show that heating of electrons occurs even with
cross-polarized laser pulses. Here we use a simple 1D model in
order to obtain an estimate of the stochastic heating for two
lasers with arbitrary linear polarizations: we neglect collective
plasma effects and follow test electrons in the laser fields. We
use gaussian laser pulses at $800\;\mbox{nm}$ with duration $\tau=
30 \,\mbox{fs}$ at FWHM with normalized strength $a_0=2$ for the
pump pulse and $a_1=0.4$ for the injection pulse, which are close
to the parameters of the experiment (after some self-focusing has
occurred). Electrons are initially randomly distributed in the
interval $z\in[-2 \tau c,\; 2 \tau c]$, $z=0$ corresponding to the
position where the two pulse maxima collide. Outside this region,
electrons interact with the laser pulses successively and we have
checked that they are not significantly heated.

\begin{figure}[!ht]
\includegraphics[width=7.5cm]{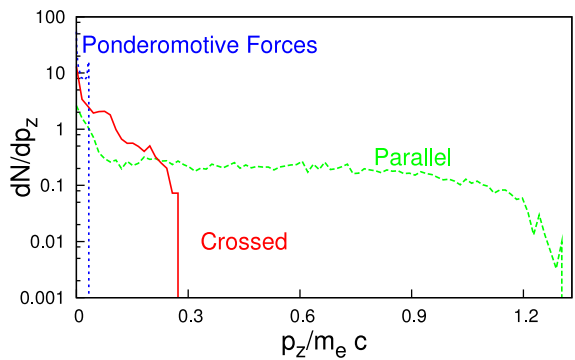}
\caption{\label{stoc} Comparison between the two heating
mechanisms: spectra after the interaction of the two pulses. Dashed line: parallel polarizations, solid line: crossed
polarizations, dotted line: ponderomotive forces only.}
\end{figure}

Figure \ref{stoc} shows the electron spectra after the collision:
the dashed line represents the heating when the
polarizations of the two pulses are parallel, the solid line
corresponds to the crossed polarizations case and the dotted
line corresponds to the case where electrons experience only the
ponderomotive forces $\nabla (<\bf{a_0^2}>+<\bf{a_1^2}>)$. The latter
case, where high frequencies are averaged, corresponds to the
reference case for which there is no stochastic nor beatwave heating.
The first remarkable feature is that electrons are more heated by
two cross-polarized pulses (solid line) than by the sum of the two
ponderomotive forces (dotted line). This can be explained by
the fact that for high laser intensities, the electron motion
becomes relativistic ($a_0>1$) which introduces a longitudinal
component $p_z$ through the $\bf{v} \times \bf{B}$ force. Thus,
the two perpendicular laser fields couple through the relativistic
longitudinal motion of electrons. This relativistic coupling makes
it possible to heat electrons.

This simple model also shows that the heating is much more
efficient when the polarizations of the two pulses are parallel
(dashed line), the maximal momentum obtained being 4-5 times
higher than in the crossed polarizations case. This is why the
injection is strongly correlated with the polarization of the two
pulses as we have seen in the experimental results.

If the electron injection were only dependent on the
preacceleration spectrum of the electrons, the crossed
polarizations case should always result in a decrease of the
trapped charge by several orders of magnitude compared to the
parallel polarizations case. This corresponds to the discussion on
polarization influence on the trapped charge made in Ref.
\cite{Fubiani}. However, in Ref. \cite{Fubiani} the
electromagnetic fields do not depend self-consistently on the
motion of electrons, and therefore the model does not take into
account an important physical phenomenon: the wakefield inhibition
\cite{poprechatin} at the collision of the laser pulses. To
exhibit the influence of this phenomenon, we have performed
self-consistent 1D PIC simulations. We have used the code CALDER
\cite{Erik} with the same laser parameters as above and an
electronic density of $n_e=7\times 10^{18}\mbox{cm}^{-3}$. The
simulation box consists of 13600 cells measuring each
$0.1/k_0\approx0.013\; \mu$m and containing $20$ pseudo-particles
each. The time step of the computation lasts $0.05/(c k_0)=2.1\;
10^{-2}\; \mbox{fs}$.

\begin{figure}[!ht]
\includegraphics[width=9cm]{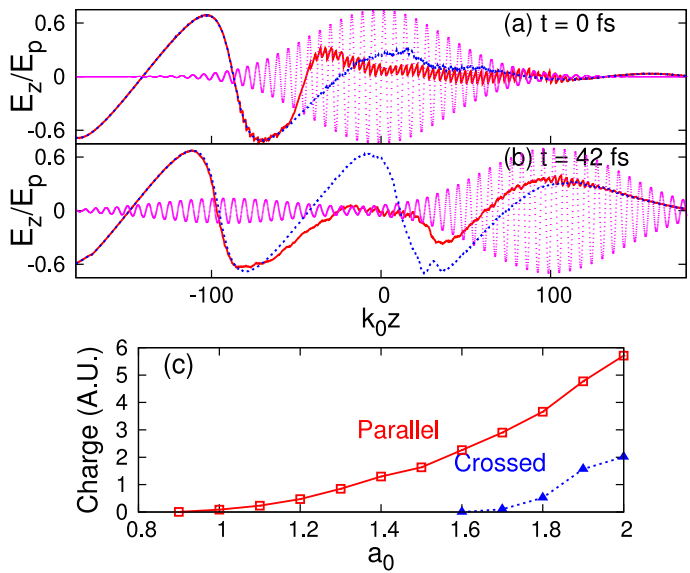}
\caption{\label{wkdes} (a) and (b): 1D PIC simulation snapshots of
the longitudinal electric field during the collision at different
times for $a_0=2$. Solid line: parallel polarizations. Dotted
line: crossed polarizations. Thin dotted line: laser fields. (c)
Injected charge vs $a_0$ (all other parameters being the same) in
1D PIC simulations. Solid square line: parallel polarizations.
Dotted triangle line: crossed polarizations.}
\end{figure}

Figures \ref{wkdes}(a) and (b) show snapshots of the longitudinal
electric field, during and after collision. The time $t=0$
corresponds to the collision of the laser pulse maxima. The solid
line corresponds to the case of parallel polarizations whereas the
dotted line corresponds to the case of crossed polarizations. The
laser fields are also represented by the thin dotted line.
When the pulses have the same polarization, electrons are trapped
spatially in the beatwave and can not sustain the collective
plasma oscillation. Therefore the plasma wave is strongly
distorted during and after the collision (see Fig. \ref{wkdes}a
and \ref{wkdes}b). It has been shown in \cite{poprechatin} that
this process has a dramatic influence on the injected charge: in
an inhibited plasma wave, it is harder to trap electrons and the
charge is reduced by approximately one order of magnitude compared
to trapping in an unaffected plasma wave.

When the polarizations are crossed, the motion of electrons is
only perturbatively disturbed compared to the motion under only one
laser, and the plasma wave is almost unaffected during the
collision, see Fig. \ref{wkdes}a and b. This tends to facilitate
trapping.

In order to scan different injection conditions we have performed
a numerical scan on $a_0$, all other parameters of the 1D PIC
simulations being the same. Figure \ref{wkdes}(c) represents the
trapped charge: it shows that the threshold for injection is
higher when polarizations are crossed. This is consistent with the
fact that electron heating is less efficient with cross-polarized
lasers. It also shows that above this threshold, there is only
approximately 3 times less charge injected with cross-polarized
pulses than in the parallel polarizations case: the charge
difference due to the heating efficiency is balanced by the charge
reduction due to wake inhibition occurring only in the parallel
polarizations case.

In these simulations, we have also checked that all trapped
electrons witness the overlapping of the two lasers, i.e.
electrons located outside the collision region are not trapped.
This rules out the wake-wake and wake-laser collisions \cite{Cary,Cary1} as
main injection processes. This has also been confirmed in 2D PIC simulations.

Simulations can also give a partial understanding of the spectra
experimentally observed with the cross-polarized pulses. In Fig.
\ref{exp}a, in the crossed polarizations case, the spectrum
exhibits a broad component at lower energy, even though the
threshold for the crossed polarizations injection is not reached.
This could be explained by a staged acceleration mechanism: the
electrons are first heated under the collision of the two lasers
but as they have not gained enough momentum to be trapped, they
slip back in the following buckets of the wakefield. They are then
further accelerated in the collision of the two wakefields until
they become trapped in the accelerating structure. In that case,
injection is not localized as it occurs in multiple buckets, and
it leads to the observed broad spectrum at low energy.
On the other hand, when operating above the threshold, the volume where the electrons reach the required energy for trapping is very localized. Therefore injection results in a narrow energy distribution beam accelerated in the first plasma wave bucket. Due to the higher threshold, this injection volume is even smaller in the crossed polarizations case than in the parallel polarizations case.  Therefore, for a given collision position, injection with crossed polarizations results in a lower charge and a smaller energy spread electron beam (see Fig. \ref{exp}b). Finally, the energy
difference (10 MeV) observed between the two cases (see Fig.
\ref{exp}b) could be explained by beamloading. The wakefield is
less distorted by the lower charge injected in the crossed
polarizations case and in consequence, electrons are accelerated
to higher energies.

In conclusion, we have demonstrated both experimentally and
theoretically a new regime of optical injection using the stochastic heating at the collision of two cross-polarized laser beams. This scheme can provide stable
monoenergetic electron beams. The injection threshold for crossed polarizations is higher and in our experiment, we found that the beams produced in this case had lower charge and smaller energy spread than those injected by the beatwave scheme. Injection with crossed polarizations is also safer for the laser system when operating in a collinear colliding geometry: two polarizers are sufficient to protect the system from laser feedbacks. This would particularly interesting for experiments using waveguides \cite{Capillary}, for which the collinearity of the laser beams is mandatory.

We acknowledge the support of the European
Community Research Infrastructure Activity under the FP6
Structuring the European Research Area program (CARE, contract
number RII3-CT-2003-506395 and EUROLEAP, contract number 028514).
X.D. and E.L. acknowledge discussions with A. Bourdier on stochastic heating.
\bibliographystyle{unsrt}

\end{document}